\documentclass[structabstract]{aa}  
\usepackage{graphicx}
\usepackage{amsmath}
\usepackage{txfonts}
\begin{document}
   \title{Exploring the long-term variability and evolutionary stage of the interacting binary DQ Velorum}

   \author{D. Barr\'ia \inst{1}, R. E. Mennickent \inst{1},  D. Graczyk \inst{1}, Z. Ko\l{}aczkowski \inst{2}}

 \institute{Universidad de Concepci\'on, Departamento de Astronom\'ia, Concepci\'on, Chile\\
              \email{dbarria@astro-udec.cl,rmennick@astro-udec.cl}              
\and
             Instytut Astronomiczny Uniwersytetu Wroclawskiego, Wroc\l{}aw, Poland \\
             \email{kolaczk@astro1.astro.uni.wroc.pl}
             }

   \date{\today}

  \abstract
  {}
   {To progress in the comprehension of the double periodic variable (DPV) phenomenon, we analyse a series of optical spectra 
   of the DPV system DQ Velorum during much of its long-term cycle. In addition, we investigate the evolutionary
   history of DQ Vel using theoretical evolutionary models to obtain the best representation for the current observed stellar and orbital
   parameters of the binary. We investigate the evolution of DQ Vel through theoretical evolutionary models to estimate the age and the mass
   transfer rate which are compared with those of its twin V393 Scorpii.}
   {We subtract the donor star contribution from the composite spectra of DQ Vel using a synthetic spectrum as a donor
   template. Donor subtracted spectra covering around 60\% of the long-term cycle, allow us to investigate time-modulated spectral
   variations of the gainer star plus the disc. We use Gaussian fits to measure the equivalent widths (EWs) of Balmer and helium lines
   in the separated spectra during the long-term cycle and thus analyse EW variabilities. We compare the observed stellar parameters of
   the system with a grid of theoretical evolutionary tracks computed under a \emph{conservative} and a \emph{non-conservative} evolution regime.}
   {We have found that the EW of Balmer and helium lines in the donor subtracted spectra are modulated with the long-term
   cycle. We observe a strenghtening in the EWs in all analysed spectral features at the minimum of the long-term cycle which might be
   related to an extra line emission during the maximum of the long-term variability.
   Difference spectra obtained at the secondary eclipse support this scenario.\\
   We have found that a non-conservative evolutionary model where DQ Vel has lost mass at some stage of its binary history, is a better
   representation for the current observed properties of the system. The best evolutionary model suggests that DQ Vel has an age
   of $7.40\times10^{7}\,\mathrm{yr}$ and is currently in a low mass transfer rate ($-9.8\times10^{-9}\,\mathrm{M_{\odot}/yr}$) stage,
   after a mass transfer burst episode. Comparing the evolutionary stages of DQ Vel
   and V393 Sco we observed that the former is an older system with a lower mass transfer rate. This might explain 
   the differences observed in the physical parameters of their accretion discs.
   }
   {}

   \titlerunning{Exploring the long-term variability and the evolutionary stage of DQ Velorum}
  \authorrunning {Barr\'ia et al.}
   \maketitle
%

\section{Introduction}

DQ Velorum is an eclipsing binary (V$\sim10.7$) with an orbital period of 6.08 $\mathrm{d}$, which belongs to the group of 
\emph{double periodic variables\/} (DPVs). This group of interacting binaries was discovered after a photometric search for Be stars in the
OGLE-II database. They are characterised by a defined relation between the orbital period ($P_\mathrm{o}$) and a second observed long-term 
periodicity ($P_\mathrm{l}$) such that $P_\mathrm{l}=\eta \times P_\mathrm{o}$ (Mennickent et al. \cite{mennickent2003}, 
Mennickent \& Ko\l{}aczkowski \cite{mennickent2010a}, Poleski et al. \cite{poleski}). 
At present, more than 200 DPVs have been discovered in the Milky Way and the Magellanic Clouds. The orbital and long-term periods calculated
for the 13 Galactic DPVs known so far, allowed to find $\eta=32.7$ for the Milky Way DPVs (Mennickent et al. \cite{mennickent2012a}).
In the current scenario, the long-term periodicity is interpreted as cycles of mass loss from the binary into the interstellar medium. 
After a deep spectroscopic analysis of the DPV system V393 Sco, Mennickent et al. (\cite{mennickent2012b}) find that 
the long-term variability could be explained in terms of an anisotropic wind of variable strength whose emisivity is modulated with the long cycle. 
However, the mechanism(s) involved in the wind strength variability are still unclear.\\
Using ASAS \footnote{http://www.astrouw.edu.pl/asas/. For further details read Pojmanski, G. (\cite{pojmanski}).} photometric data of DQ Vel, 
Michalska et al. (\cite{michalska}) detected an additional long photometric variability of 188.9 $\mathrm{d}$ 
which agrees with the previously mentioned $P_\mathrm{l}/P_\mathrm{o}$ DPV relation. Using the data obtained after a photometric and spectroscopic 
campaign carried out during 2008-2011, Barr\'ia et al. ({\cite{barria}}) determined the fundamental stellar and orbital parameters of this binary. 
Their techniques involved the separation of the DQ Vel composite spectra, radial velocity (RV) measurements, and a multicomponent fit to the V-band
ASAS light curve.
The authors argued that DQ Vel is a semi-detached system consisting of a more massive B3V primary star (hereafter gainer) and an A1III secondary
(hereafter donor) star plus an extended accretion disc around the gainer. 
They obtained a spectroscopic mass ratio $q=0.31\pm0.03$, masses of $M_\mathrm{d}=2.2\pm0.2\ M_{\odot}$ and $M_\mathrm{g}=7.3\pm0.3\ M_{\odot}$, 
the radii $R_\mathrm{d}=8.4\pm0.2\ R_{\odot}$ and $R_\mathrm{g}=3.6\pm0.2\ R_{\odot}$, and temperatures of $T_\mathrm{d}=9400\pm100\ \mathrm{K}$ and
$T_{g}=18\,500\pm500\ \mathrm{K}$ for the stellar components. Also, the physical and geometric parameters for the
accretion disc were obtained. Interestingly, strong similarities were found in the stellar parameters of DQ Vel and the well-studied
DPV system V393 Sco (Mennickent et al. \cite{mennickent2010b}, \cite{mennickent2012a}, \cite{mennickent2012b}). However, the two systems show different
geometric and physical properties in their accretion discs. This latter might be related to different evolutionary stages.\\
The growing collection of discovered DPVs suggests that they represent a significant stage in the lifetime of a binary. Around 30\% of them are 
eclipsing and only 2\% have been studied spectroscopically (V393 Sco, DQ Vel, OGLE05155332-6925581, AU Mon (in preparation) and 
HD170582 (in preparation)). 
In this sense, it becomes important to extend the number of well studied systems to investigate the nature and evolution of DPVs.
To do this, main physical properties of the binary components together with spectroscopic analysis during the long-term cycle are strongly required. \\
In this contribution we present the first spectroscopic analysis of DQ Vel during much of its long photometric cycle. Our analysis includes composite and
separated spectra in different orbital and long phases. We also investigate the evolutionary history of DQ Vel and compare this results with evolutionary 
studies of V393 Sco.\\

\section{Data Analysis}
In this analysis we use the sample of 46 optical spectra previously studied by Barr\'ia et al. (\cite{barria}), supplemented by six additional
spectra collected during the last years. A detailed description and reduction processes for the former spectra are given in Table 1 and Section 2
on the aforementioned paper. The six additional optical spectra were obtained at different instruments/telescopes:
three of them were collected during three nights on June-2012 with the echelle spectrograph at the $2.5\,\mathrm{m}$ \emph{Ir\'{e}n\'{e}e du Pont} 
telescope located at the Las Campanas Observatory, two spectra were obtained with the echelle spectrograph FEROS mounted at the $2.2\,\mathrm{m}$
ESO/MPG telescope at La Silla and, one spectrum was collected with the echelle spectrograph MIKE mounted at the $6.5\,\mathrm{m}$ Clay telescope in
Las Campanas Observatory. A summary of these spectra is given in Table 1. The orbital and long cycle phases were calculated, according to
the following ephemerides (Barr\'ia et al. \cite{barria}):

\begin{equation}
   \begin{array}{l}
  T_\mathrm{min,orb}= HJD\,2453407.60(2)+6.083299(7)\times E \\
  T_\mathrm{max,long}= HJD\,2453437.20(16)+188.7(2)\times E\,.
   \end{array}
   \end{equation}
 \\
 In Figure 1 we show the histograms for the orbital and long cycle phases covered by the total spectra analysed in this paper. 
 Our observations cover 90\% of the orbital variability and 60\% of the long-term periodicity.
\begin{figure}
   \centering
   \includegraphics[width=8.5cm]{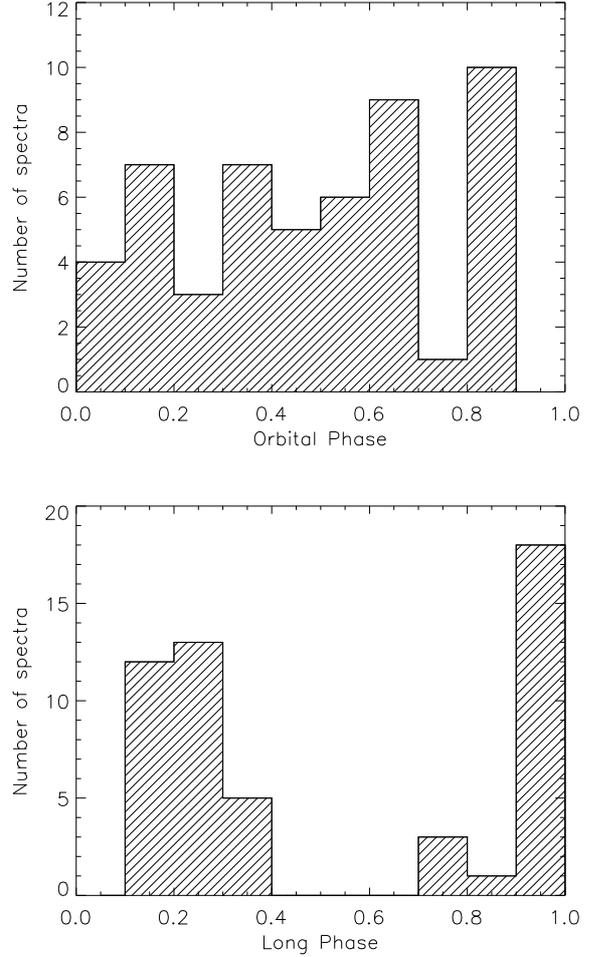}
   \caption{\textit{Top panel}: Number of spectra per bins of 0.1 orbital phases. \textit{Bottom panel}: Number of spectra per bins of 0.1 long phases.}
   \label{Fig1}
    \end{figure}
    \\
\begin{table*}
\caption{Details of the new spectra included in this study. $\phi_{o}$ is the orbital phase and  $\phi_{l}$ corresponds to the long cycle phase.}     
\centering          
\begin{tabular}{c c c c c c c c c} 
\hline\hline       
Observatory & Telescope/Instrument & UT date & HJD-2450000 & $\phi_{o}$ & $\phi_{l}$ & Exp time(s) & S/N & wavelength\\ 
            &                      &         &             &            &            &             &     & coverage ($\mathrm{\AA{}}$)\\
\hline                    
 Las Campanas & Du Pont/echelle & 2012-06-24 & 6103.473  & 0.16 & 0.13 & 120 & 65 & 3630-5260 \\  
 Las Campanas & Du Pont/echelle & 2012-06-25 & 6104.448  & 0.32 & 0.14 & 120 & 77 & 3630-5260 \\     
 Las Campanas & Du Pont/echelle & 2012-06-26 & 6105.461  & 0.48 & 0.14 & 120 & 65 & 3630-5260 \\  
 La Silla     & MPG/FEROS       & 2012-05-10 & 6057.662  & 0.63 & 0.89 & 600 & 50 & 3530-9220\\  
 La Silla     & MPG/FEROS       & 2013-02-19 & 6342.652  & 0.48 & 0.40 & 1500 & 55 & 3530-8850\\ 
 Las Campanas & Clay/MIKE       & 2011-04-05 & 5657.483  & 0.84 & 0.77 & 600 & 47 & 3450-5050 (b)\\
              &                 &            &           &      &      &     &    & 4850-9150 (r) \\
\hline     
\end{tabular}
\tablefoot{The S/N ratio was measured at the continuum level using the ranges 4500-4800 \& 5000-5250 $\mathrm{\AA{}}$ for Du Pont spectra; 
4500-4800 \& 5100-5800 $\mathrm{\AA{}}$ for MIKE and FEROS spectra.}             
\end{table*}
The Du Pont/echelle and the MPG/FEROS spectra listed in Table 1, were subjected to the same reduction processes described in Barr\'ia et al. 
(\cite{barria}). The spectral range for the Du Pont/echelle spectra is 3630-5260 $\mathrm{\AA{}}$. The MPG/FEROS spectra has a wavelength coverage
from 3530 $\mathrm{\AA{}}$ to around 9000 $\mathrm{\AA{}}$.\\
The MIKE instrument is a double echelle spectrograph comprised of two arms (red and blue) with independent CCDs. The standard wavelength coverage for 
the blue arm is around 3350-5000 $\mathrm{\AA{}}$ and 4900-9500 $\mathrm{\AA{}}$ for the red arm. The instrument resolutions (for one arc-second slit) are
$28\,000$ and $22\,000$ for blue and red side respectively. An internal Thorium Argon lamp is used for wavelength calibration.
The standard reduction processes like flat and bias correction, wavelength calibration, continuum normalization, order merging and cosmic rays subtraction 
were applied using IRAF \footnote {IRAF (http://iraf.noao.edu/) is distributed by the National Optical Astronomy Observatories and operated by the Association
of Universities for Research in Astronomy Inc., under cooperative agreement with the National Science Foundation.} tools. 
The reduction processes were applied separately in the blue and red side of the MIKE spectrum. 
Heliocentric corrections were applied to all spectra listed in Table 1.

\section{Results and Discussion}
\subsection {Long cycle spectroscopic analysis}
Donor and gainer radial velocities (RVs) were measured for the spectra listed in Table 1. To compute the donor RVs, we use a donor
template spectrum and apply a \textit{cross correlation process} (CCR) using IRAF \textit{fxcor} task in the 4350-5000 $\mathrm{\AA{}}$ wavelength
range where we find more lines coming from the donor. We select from a grid of synthetic spectra, a donor template spectrum with
$T_\mathrm{d}= 9400\ \mathrm{K}$, $\log g_\mathrm{d}$ = 2.9, $v_\mathrm{r}\ \sin i = 75 \mathrm{km\,s^{-1}}$ and solar metallicity, according to
the donor parameters found by Barr\'ia et al. (\cite{barria}). The grid of synthetic spectra was computed using atmospheric models with
the line-blanketed LTE ATLAS9 code (Kurucz \cite{kurucz}), which treats line opacity with opacity distribution functions (ODFs). The Kurucz models are
constructed assuming plane-parallel geometry and hydrostatic and radiative equilibrium of the gas. The synthetic spectra were computed with the SYNTHE
code (Kurucz \cite{kurucz}). Both codes, ATLAS9 and SYNTHE, were ported under GNU Linux by Sbordone (\cite{sbordone}) and are available online \footnote{wwwuser.oat.ts.astro.it/atmos/}.
The atomic data were taken from Castelli \& Hubrig \cite{castelli} \footnote{http://wwwuser.oat.ts.astro.it/castelli/grids.html}. The theoretical 
models were obtained for effective temperatures from 6000 to 19\,000 $\mathrm{K}$ with steps of 100 $\mathrm{K}$ and for surface gravities from 2.0 to 
4.5 dex with steps of 0.1 dex. Solar and 0.5 dex higher metallicities were taken into account. The grid of synthetic spectra was calculated for
different rotation velocities, $v_\mathrm{r} \sin i =$ 0, 25, 50, 75, 100 and 120 $\mathrm{km\,s^{-1}}$.\\
We observe that the He\,I absorption lines move in anti-phase with the donor RVs, which agrees with our previous work. This fact suggests that these 
lines are originated in or close to the gainer star. Due to the short wavelength coverage of the Du Pont/echelle spectra (3630-5260 $\mathrm{\AA{}}$),
the gainer RVs are calculated using gaussian fits to the He\,I 4026.191/4387.929/4471.479 $\mathrm{\AA{}}$ lines and then computing their Doppler
 shifts. The gainer RVs for the individual spectra are obtained taking the mean value of the velocities calculated for each of 
 the He\,I lines.  
Final RVs for the donor ($RV_{d}$) and gainer ($RV_{g}$) are listed in Table 2. 
\begin{table}
\caption{Calculated RVs for the spectra listed in Table 1.}     
\centering          
\begin{tabular}{c c c r} 
\hline\hline       
 HJD-2450000 & $\phi_{o}$ & $RV_{d}\,(\mathrm{km\,s^{-1})}$ & $RV_{g}\,(\mathrm{km\,s^{-1})}$ \\ 
\hline                    
  6103.473  & 0.160 & $98.17\pm0.88$ &  $-77.86\pm14.92$ \\  
  6104.448  & 0.320 & $131.09\pm2.80$ & $-47.75\pm8.45$  \\     
  6105.461  & 0.486 & $18.34\pm0.06$ &  $-79.11\pm5.20$ \\  
  6057.662  & 0.630 & $-121.63\pm0.93$ & $9.63\pm3.16$  \\  
  6342.652  & 0.480 & $5.27\pm0.11$    & $-24.24\pm10.58$\\
  5657.483  & 0.846 & $-137.97\pm1.55$ &  $34.92\pm7.41$ \\ 
\hline     
\end{tabular}
\end{table}
\\
Remarkable variabilities in the spectra of DQ Vel are observed along the orbital and long-term phases. To analyse the long-term behaviour of the gainer
star and the disc, we subtract from each spectra the donor template spectrum multiplied by a light contribution factor obtained from the light curve model. 
These factors depends on the orbital phase and are calculated for a specific wavelength range around a spectral line 
(for further details see Barr\'ia et al. \cite{barria}, Section 3.2).
\begin{figure}
   \centering
   \includegraphics[width=9.3cm]{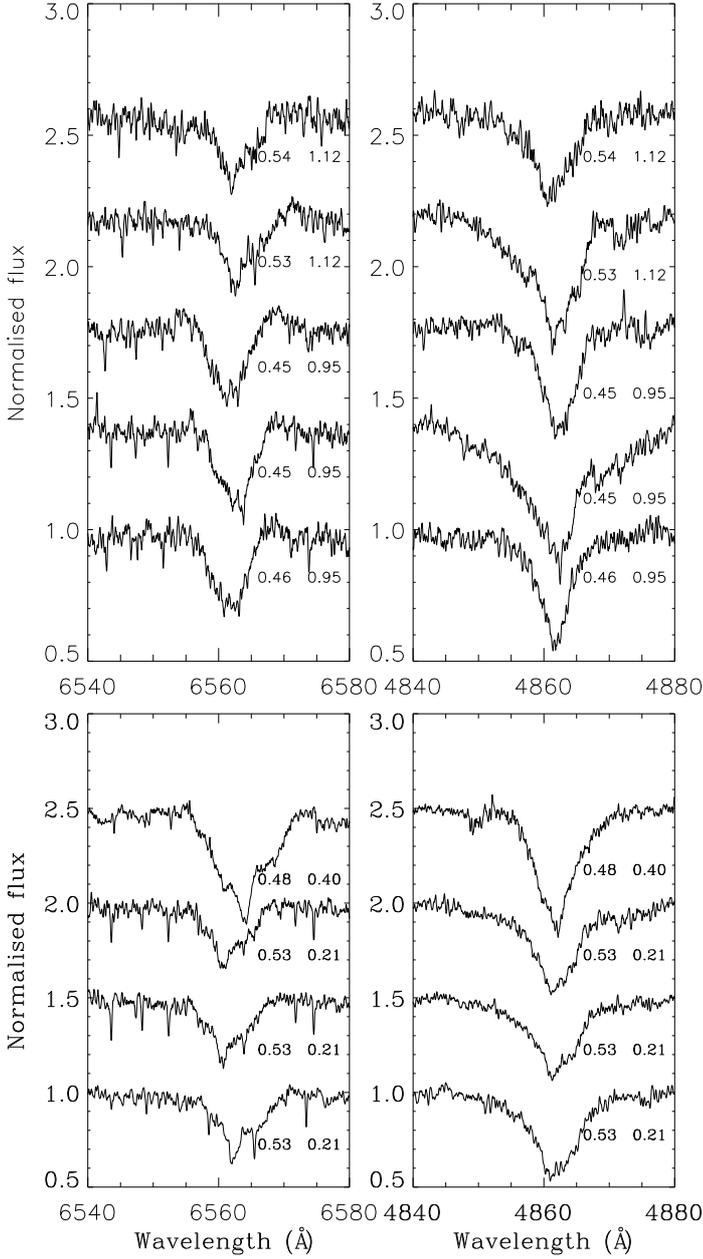}
   \caption{Donor-subtracted spectra of DQ Vel in the $H\alpha$ (left panels) and $H\beta$ (right panels) regions around the secondary eclipse. Spectra are shown at the
   maximum of the long-term cycle (upper panels) and on the way to the minimum of the long-term variability (lower panels). Labels indicate, orbital (left) and long cycle (right) phases.
   The spectra were offset for clarity.}
   \label{Fig2}
    \end{figure}
\begin{figure}
   \centering
   \includegraphics[width=9.3cm]{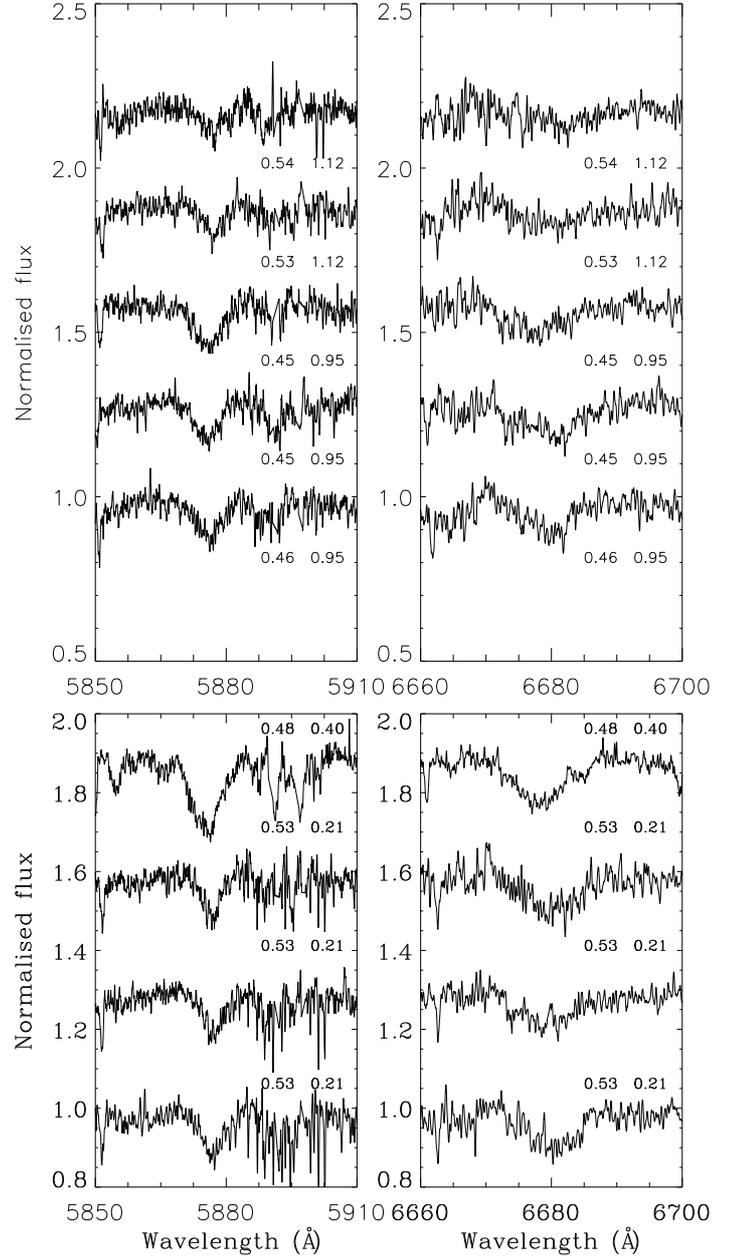}
   \caption{Donor-subtracted spectra of DQ Vel in the He\,I$\,5875$ (left panel), and He\,I$\,6678$ (right panel) regions around the secondary eclipse.
   Labels indicate, orbital (left) and long cycle (right) phases.}
   \label{Fig3}
    \end{figure}
    \\
In Figures 2 and 3 we show selected donor-subtracted spectra in the H$\alpha$, H$\beta$, He\,I\,5875 $\mathrm{\AA{}}$ and He\,I\,6678 
$\mathrm{\AA{}}$ regions around the mid eclipse $\phi_{o}\sim 0.5$. At the upper panels, spectra along the maximum of the long-term cycle are shown.
Here, weak double-peaked emission of variable strenght are seen in the H$\alpha$ profiles. These emissions can reach velocities up
to $\pm350\,\mathrm{km\,s^{-1}}$. Just after the maximum of the long-term variability ($\phi_{l}\sim 1.1$), we observe a weakeness
of the central absorption in all profiles. It is remarkable the case of the He\,I\,6678 $\mathrm{\AA{}}$ line, which seems to dissapear at this long-term phase,
as can be seen at the upper right panel in Figure 3. 
Asymmetric H$\alpha$ and H$\beta$ profiles are observed at different long-term phases (see for instance the spectra at $\phi_{l}=0.95,1.12$ in Figure 3).
The origin of these features is unkwon but might be related to an asymmetric distribution of the circumstellar matter at different epochs.\\
In the lower panels of Figures 2 and 3, we show spectra toward the
minimum of the long-term variability ($0.2 <\phi_{l} < 0.4$). We observe a strenghtening of the central absorption at $\phi_{l}=0.4$ in all profiles. 
We measure the equivalenth width (hereafter EW) of the central absorption for the spectral lines displayed in Figures 2 and 3 along the long-term cycle. 
On average, higher EWs values are found during the minimum of the long-term cycle as can be seen in Figure 4. To check if this pattern is a general trend, 
we measured the EWs of H$\alpha$, H$\beta$, He\,I\,5875 $\mathrm{\AA{}}$ and He\,I\,6678 $\mathrm{\AA{}}$ lines for all donor subtracted spectra during the maximum and 
the minimum of the long-term variability. As can be seen in Figure 5, during the minimum of the long-term we observe an increase in the EWs, more significant at the
main eclipse ($\phi_{o}\sim 0.0$) when the gainer and the disc are hidden by the donor star. A similar pattern is also detected in H$\gamma$ and H$\delta$ profiles, as 
well as in other helium lines.
An extra global emission such that suggested by Mennickent et al. (\cite{mennickent2012a}) during the high state of the long cycle could explain this 
variability pattern.
\begin{figure}
   \centering
   \includegraphics[width=9.3cm]{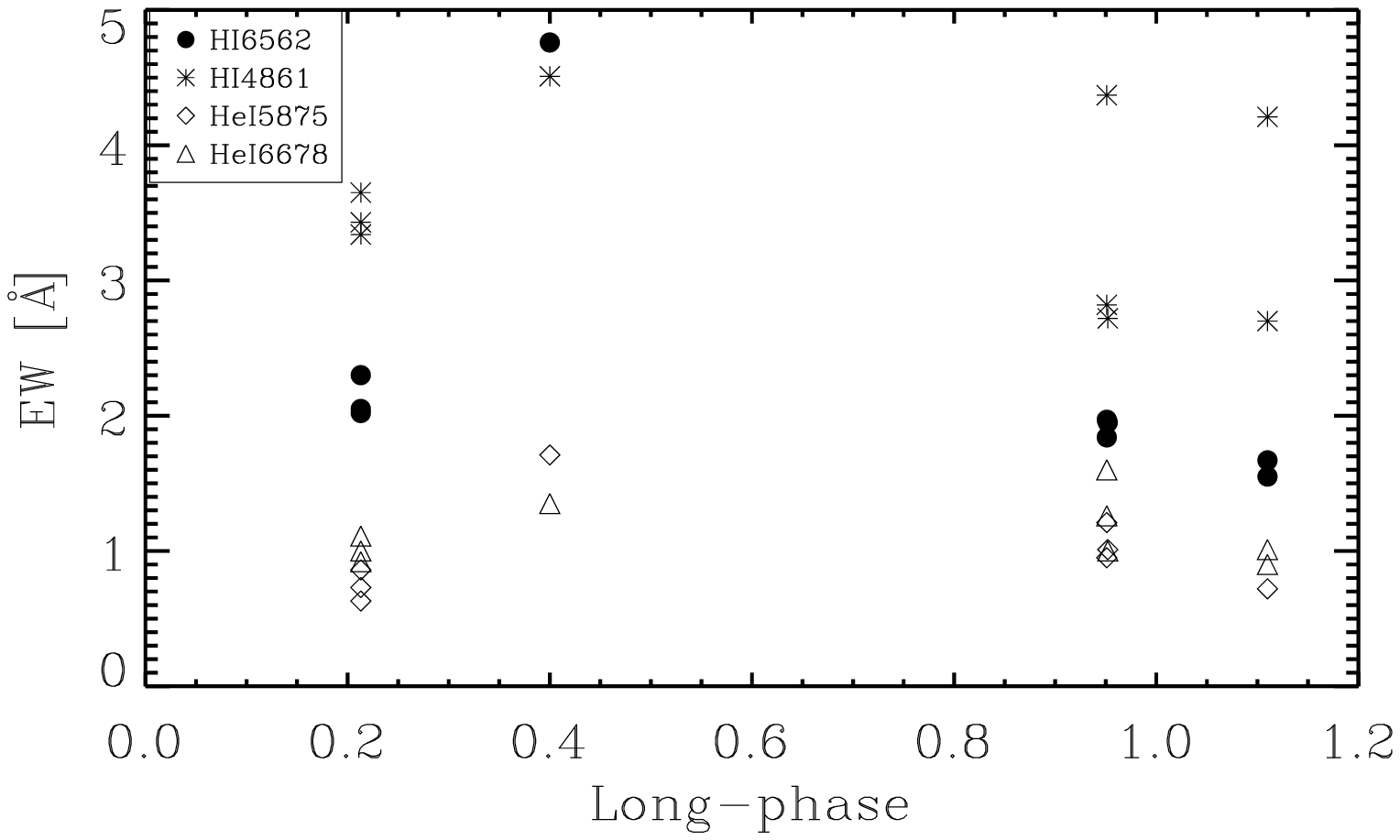}
   \caption{Equivalent width of the H$\alpha$, H$\beta$, He\,I\,5875 $\mathrm{\AA{}}$ and He\,I\,6678 $\mathrm{\AA{}}$ lines along the long-term cycle and around the secondary
   eclipse.}
   \label{Fig4}
    \end{figure}
    \begin{figure}
   \centering
   \includegraphics[width=8.5cm]{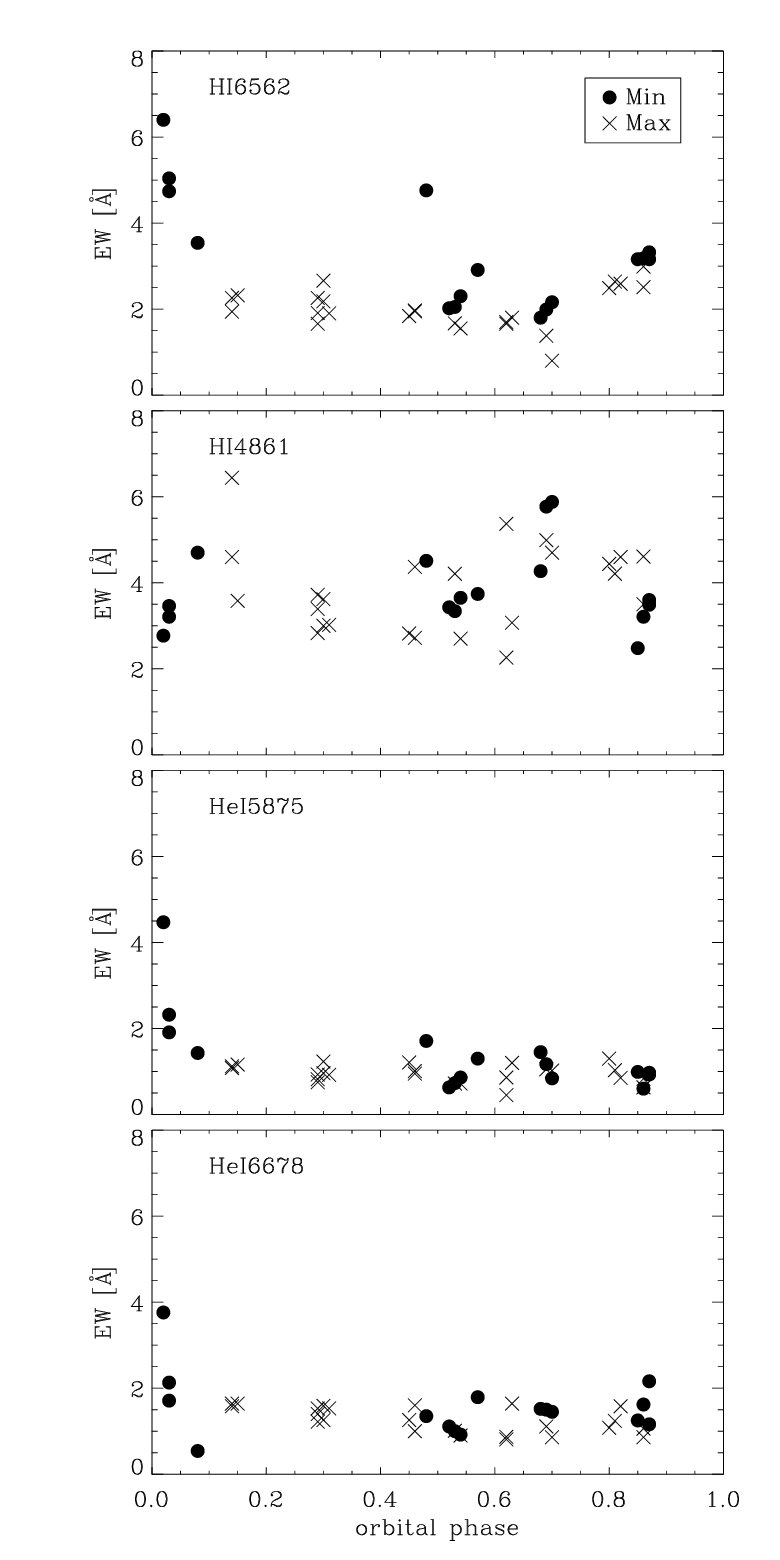}
   \caption{Equivalent width of the H$\alpha$, H$\beta$, He\,I\,5875 $\mathrm{\AA{}}$ and He\,I\,6678 $\mathrm{\AA{}}$ lines along the orbital phase at the maximum and minimum
   of the long cycle.}
   \label{Fig5}
    \end{figure}

\subsection{The O\,I\,7773 line and DACs}
Donor subtracted spectra showing the time variability of the O\,I\,7773 $\mathrm{\AA{}}$ line during the orbital and long cycle are displayed in Figure 6. 
Radial velocity measurements of the central absorption of O\,I\,7773 $\mathrm{\AA{}}$ in the composite spectra show that it is in phase with the donor star. 
The half-amplitude of the RV curve is $136.1\pm3.8 \, \mathrm{km\,s^{-1}}$ for a circular-orbit solution. This measure is around 25\% lower that the mean 
half-amplitude value for the donor RV curve found in our previous paper. This can be explained considering that the O\,I\,7773 $\mathrm{\AA{}}$ RV curve
was obtained using only five RV measurements from the spectra available at this wavelength range. Moreover, the O\,I\,7773 $\mathrm{\AA{}}$ absorption
line is broader, asymmetric and affected by enhanced blue/red absorption wings.\\ 
Following the tendency of Balmer and He lines, we observe a deeper and broader central absorption of the O\,I\,7773 $\mathrm{\AA{}}$ line at 
$\phi_{l}=0.4$ and a weakeness of this spectral feature at $\phi_{l}=0.88$. 
Blue depression wings are observed in three spectra at orbital phases $\phi_{o}=0.32, 0.48$ and $0.65$ and at different long-term phases. Here,
weak discrete absorption components (DACs) at the blue ($\phi_{o}=0.48, 0.65$) and at the red side ($\phi_{o}=0.32$) of the line have been detected.
DACs are observed as peculiar line profiles in hot OB, Oe and Be-type stars. DACs have been interpreted as absorption lines created at different density 
regions moving radially at distinct velocities (Danezis et al. \cite{danezis}). Mennickent et al. (\cite{mennickent2012b}) have interpreted DACs observed in the DPV
system V393 Sco as signatures of a clumpy wind emerging from the disc in the direction of the binary motion. Interestingly, no DACs are observed at 
$\phi_{o}=0.84$ ($\phi_{l}=0.76$) and neither blue or red enhanced absorption wings. Despite the fact that additional spectra are required to confirm
this pattern, it seems that enhanced absorption wings and DACs are correlated. A possible explanation for this, might be related to the hot spot 
(hot line), which reach its maximum visibility around $\phi_{o}=0.84$. If the DACs and the enhanced absorption wings are related to a wind, then the 
hot spot may acts veiling the wind effects, specially if the strenght of this latter is weak.
Considering that DACs are observed only at three of the five spectra available at this wavelength range, these results are similar to those observed by
Mennickent et al. (\cite{mennickent2012b}) in spectroscopic studies of V393 Sco.
\begin{figure}
   \centering
   \includegraphics[width=9cm]{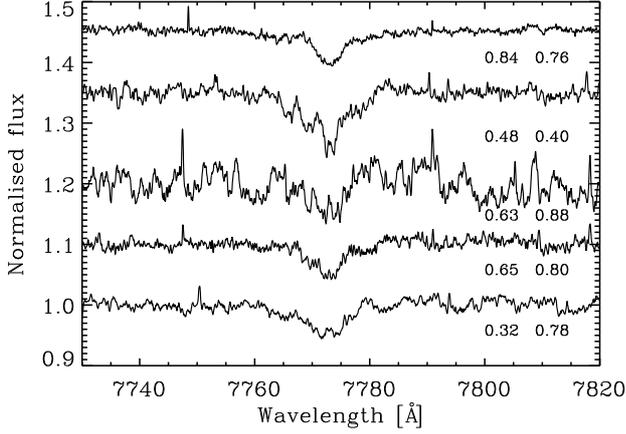}
   \caption{Donor-subtracted spectra of DQ Vel in the O\,I\,7773 $\mathrm{\AA{}}$ region. Labels indicate orbital (left) and long-term (right) phases.}
              \label{Fig6}                                    
    \end{figure}
    
\subsection{Difference Profiles}
 Donor-substracted spectra of DQ Vel along the orbital motion show that the intensity of the H$\alpha$ central absorption is 
 weakened outside the eclipses (see Figure 6, Barr\'ia et al. \cite{barria}). This fact can be related to the extra-emission 
 component coming from the interaction region between the gas stream and the accretion disc. 
 To minimize this emission contribution and thus investigate other possible extra absorption/emission features, we analyse
 a series of higher S/N spectra around the secondary eclipse. A sample of H$\alpha$ composite spectra of DQ Vel at the mid eclipse are 
 displayed in Figure 7. Here, weak emission components in the wings of the line profile are observed. However, the weakness of these emission 
 features compared to the continuum flux, inhibit any type of quantitative analysis of the circumstellar/binary material.
 Intensities or equivalent widths analysis of any extra absorption and/or emission feature in the system are not possible due to blending effects 
 with photospheric absorption. One means to analyse these emission/absorption features is by subtracting the stellar photospheric absorption profiles
 from the observed spectra to produce the so-called \textit{difference profiles} (Richards \cite{richards1993}).\\
 If the helium lines observed in DQ Vel spectra are representative of the gainer motion, then we can use the gainer stellar parameters obtained from
 the light curve fit to generate a gainer synthetic spectrum. This can be added to the donor template to create a synthetic composite
 spectra for DQ Vel. From the grid of synthetic spectra (see Section 3.1) we select a gainer spectrum with
 $T_\mathrm{d}= 18\,500\ \mathrm{K}$, $\log g_\mathrm{d}$ = 4.1, $v_\mathrm{r}\ \sin i = 120 \,\mathrm{km\,s^{-1}}$ and solar metallicity. 
 At different epochs, donor and gainer synthetic spectra were corrected by radial velocity shifts and multiplied by a light contribution factor
 before being summed.\\ 
 In Figure 8 is displayed a comparision between the observed and synthetic (donor+gainer) spectra in the blue spectral region. 
 At the left-hand panel of Figure 9 a sequence of H$\alpha$ observed and synthetic composite spectra at different long-term phases are displayed.
 The right-hand panel show their difference spectra. An irregular emission/absorption profile is observed at the minimum of the long-term 
 variability ($\phi_{l}=0.40$). At the high state ($\phi_{l}=0.95, 1.12$ and $1.21$) a regular single-peaked 
 emission seem to dominates the difference spectra.\\
 We measured the total EWs for these emission/absorption components using multiple Gaussian fits and plotted these against the long-term phase
 as is shown in Figure 10. Higher EW values are found at the long-term phases $\phi_{l}=0.12$ and $0.21$ and the lowest EW measurement is found at 
 the low state ($\phi_{l}=0.4$). This result support the EW variations found in the central absorption (hereafter CA) of Balmer and helium lines at the donor-subtracted spectra
 (see Section 3.1).
 At higher H$\alpha$ CA intensities lower emission EW values. Variations on the circumstellar material emissivity support the scenario for an
 extra-emission appearing on the system during the long-term variability. 
 \begin{figure}
   \centering
   \includegraphics[width=8cm,angle=0]{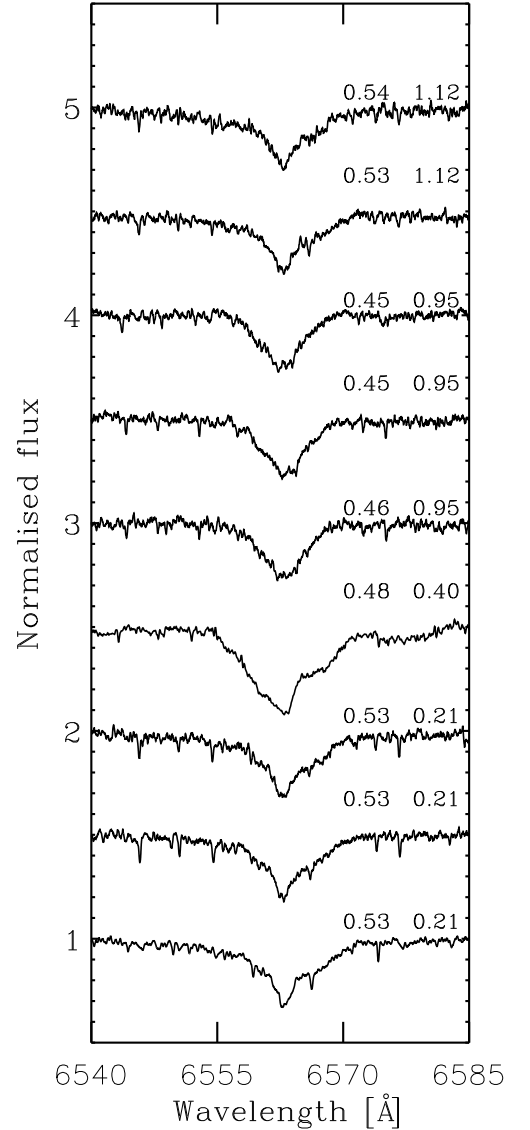}
   \caption{A sequence of H$\alpha$ composite spectra for DQ Vel around the secondary eclipse.}
              \label{Fig7}                                    
    \end{figure}
\begin{figure}
   \centering
   \includegraphics[width=9.5cm,angle=0]{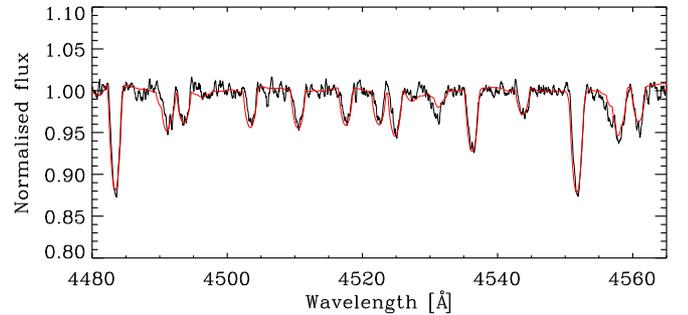}
   \caption{Comparision between the observed (black line) and the synthetic (red line) composite spectra for DQ Vel in the blue spectral region.}
              \label{Fig8}                                    
    \end{figure}
\begin{figure}
   \centering
   \includegraphics[width=9.2cm,angle=0]{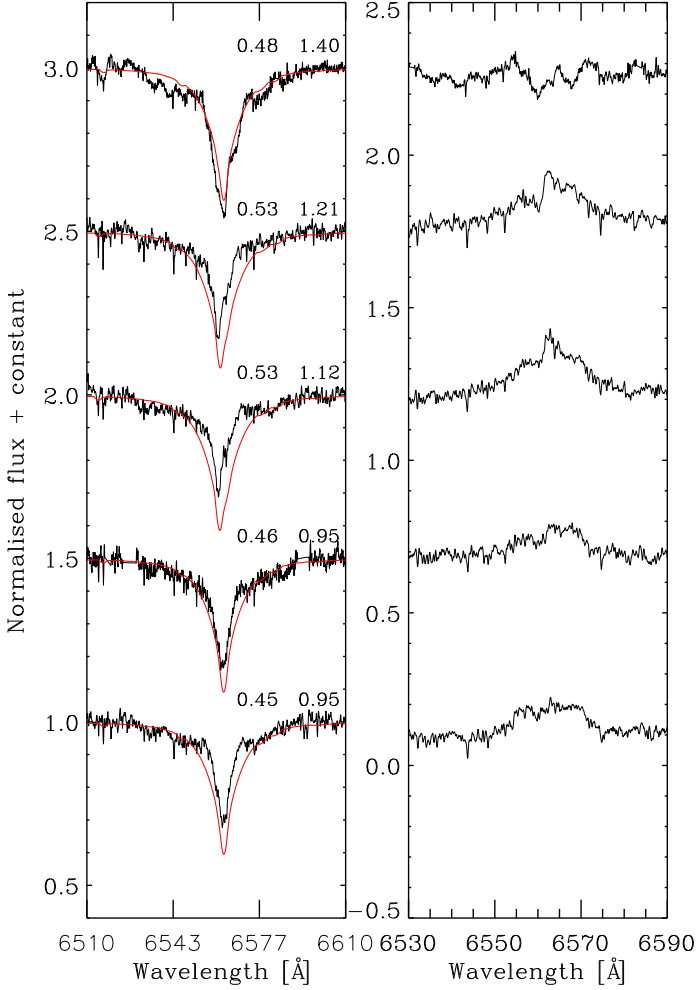}
   \caption{Left-panel: Observed and synthetic (donor+gainer) spectra of DQ Vel at different long-term phases during the secondary eclipse. The labels indicate
   the orbital (left) and long (right) phases. Right panel: H$\alpha$ difference profiles (observed - synthetic).}
              \label{Fig9}                                    
    \end{figure}
    \begin{figure}
   \centering
   \includegraphics[width=9.4cm,angle=0]{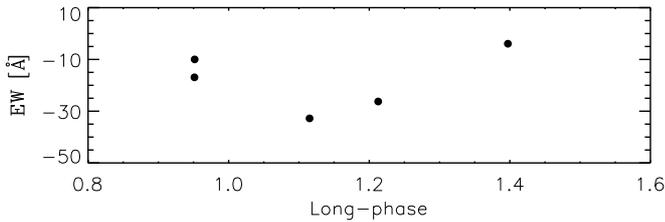}
   \caption{EWs of H$\alpha$ residual spectra displayed in Figure 9.}
              \label{Fig10}                                    
    \end{figure}

\subsection {Evolutionary stage}
The donor and gainer physical parameters can be used to analyse the stellar locations in the HR diagram. 
In Figure 11, we display the DQ Vel components together with a set of theoretical evolutionary tracks obtained from Bressan et al. 
(\cite{bressan}) for single stars with $z=0.02$ and for different stellar masses.
A first inspection shows an over-luminous donor star considering its mass of $M_\mathrm{d}=2.2\,M_{\odot}$ and $\log g_\mathrm{d}=2.9$.
This scenario can be explained if we consider that the donor is an evolved star which has swollen and cooled. Also, the large radius found for the
donor star ($R_\mathrm{d}=8.4\,R_{\odot}$) supports this assumption. The observed light curve variations outside the eclipses together 
with the donor rotational projected velocity of $v_\mathrm{d} \sin i = 69\,\mathrm{km\,s^{-1}}$ (Barr\'ia et al. \cite{barria}) 
suggest a non-spherical and synchronous rotator star which is filling its Roche lobe. \\
The gainer with a mass of $M_\mathrm{g}=7.3\,M_{\odot}$ and $\log g_\mathrm{g}=4.2$ (Barr\'ia et al. \cite{barria}) appears
as an under-luminous and cooler star. We can explain the lower luminosity and temperature values by a gainer star which is surrounded by 
circumstellar material. Under-luminous gainers has been also found in others DPVs such as AU Mon (Djurasevic et al. \cite{djurasevic}) and 
V393 Sco (Mennickent et al. \cite{mennickent2012a}).\\
We compare the physical parameters of DQ Vel with the absolute parameters of 61 semidetached Algol-type binaries analysed by 
Ibano\v{g}lu et al. (\cite{ibanoglu2006}). Two binary systems in the sample (BM Ori and RY Per) show similar 
properties to those of DQ Vel such as the orbital period, mass ratio, radii and temperatures. However, the stellar components of DQ Vel are 
more massive than the donor and gainer star of both Algol systems.
A comparision of the absolute parameters of DQ Vel with those of the total semidetached systems analysed by the authors, shows 
that DQ Vel follow the general tendency of classical semidetached Algol-type with an under-luminous 
gainer located in the main sequence band and an over-luminous donor evolved off from the main sequence with respect to 
single ZAMS and TAMS main sequence stars. However, after compare DQ Vel with the total semidetached Algol systems in the planes 
$\log\,M-\log\,R$, $\log\,M-\log\,T$, 
$\log\,R-\log\,T$, $\log\,M-\log\,L$ and $\log\,P_{o}-\log\,L$ (see Figures 3 and 4 in Ibano\v{g}lu et al. \cite{ibanoglu2006}), we found that
the DQ Vel components are located at the upper limit regions in all diagrams with more luminous, more massive and
hotter stars than the average values given for the Algols sample.
\begin{figure}
   \centering
   \includegraphics[width=6.2cm,angle=-90]{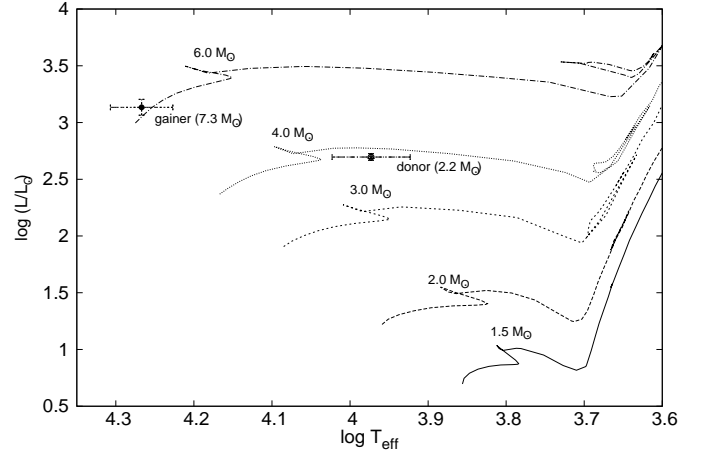}
   \caption{Theoretical evolutionary tracks for single stars together with the DQ Vel donor and gainer locations in the HR diagram. Labels indicate
   different stellar masses.}
              \label{Fig11}
    \end{figure}
\\
The evolutionary history of DQ Vel can be predicted from a comparision of the current stellar parameters with binary evolutionary models. A catalog of 561 
evolutionary tracks for binaries which includes mass loss at some stage of the binary history were computed by Van Rensbergen et al. (\cite{vanrensbergen2011})
and are available at the Centre de Donn\'ees Stellaires (CDS). This set of evolutionary tracks were calculated considering a conservative binary 
evolution (no mass loss from the system) and, a \textit{liberal} scenario where mass and angular momentum are loss during episodes of rapid mass 
transfer after the onset of RLOF. In the former, the gainer star is rotating synchronously ($f_{g}=1.0$) 
and, in the latter, the gainer has spin up and its rotation is non-synchronous ($f_{g}=0.1$). For the \textit{liberal} case, each model was computed under 
a weak and a strong tidal interaction regime. \\
In Barr\'ia et al. (\cite{barria}) the stellar parameters of DQ Vel were obtained by modeling the V-band light curve under two scenarios: a gainer 
under a synchronous rotation regime ($f_\mathrm{g}=1.0$) and a gainer star in a critical rotation scheme where $f_\mathrm{g}=12.8$. At that time, 
no significant differences (inside the errors) where found for the two fitting solutions. Here, through a multi-parametric fit, we compare each 
theoretical evolutionary model with the stellar and orbital parameters 
obtained for a synchronous and a critical rotator gainer. The observed parameter such as masses, radii, temperature, luminosities and orbital period
are compared with the theoretical values, while the mass loss rate, Roche lobe radii, chemical composition, fraction of accreted mass and age, are free-parameters.
To discriminate between the different solutions to the fit, we calculate the minimum value for the quantity $\chi_{i,j}^{2}$ which is defined by
 \begin{equation}
  \chi_{i,j}^{2}= \frac{1}{N}\varSigma_{k}w_{k}[(S_{i,j,k}-O_{k})/O_{k}]^{2},
 \end{equation}
where, $O_{k}$ is the observed \textit{k}th stellar parameter, $S_{i,j,k}$ represent the \textit{i}th synthetic model at the time \textit{t$_{j}$}, N is a 
normalization factor and, \textit{w$_{k}$} a statistical weight related to the error associated to the observed parameter $O_{k}$
(Mennickent et al. \cite{mennickent2012a}).
The minimum value for $\chi_{i,j}^{2}$ corresponds to the best evolutionary model for DQ Vel. In Figure 12, we show the best 
match (lower values for $\chi_{i,j}^{2}$) for the evolution of DQ Vel with a gainer star under a synchronous and a critical rotation scenario.
The crosses conected by a line represent the time evolution of the donor mass and the circles joined by a line are the mass transfer rate. 
The vertical dashed
line represents the current stage of DQ Vel. In Table 3, we list the best theoretical stellar and orbital parameters obtained for each case.
    \begin{figure}
   \centering
   \includegraphics[width=9.5cm]{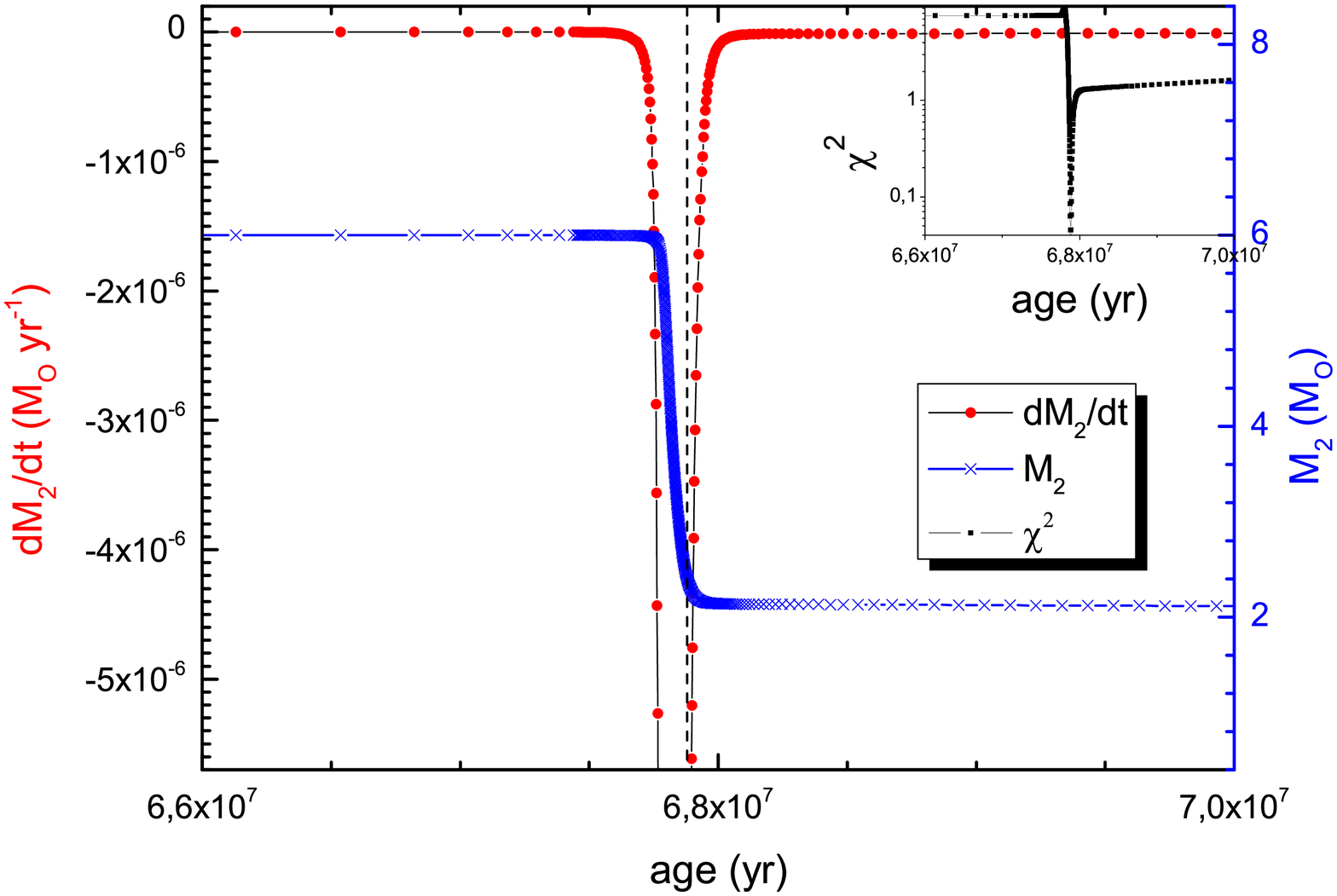}
   \includegraphics[width=9.2cm]{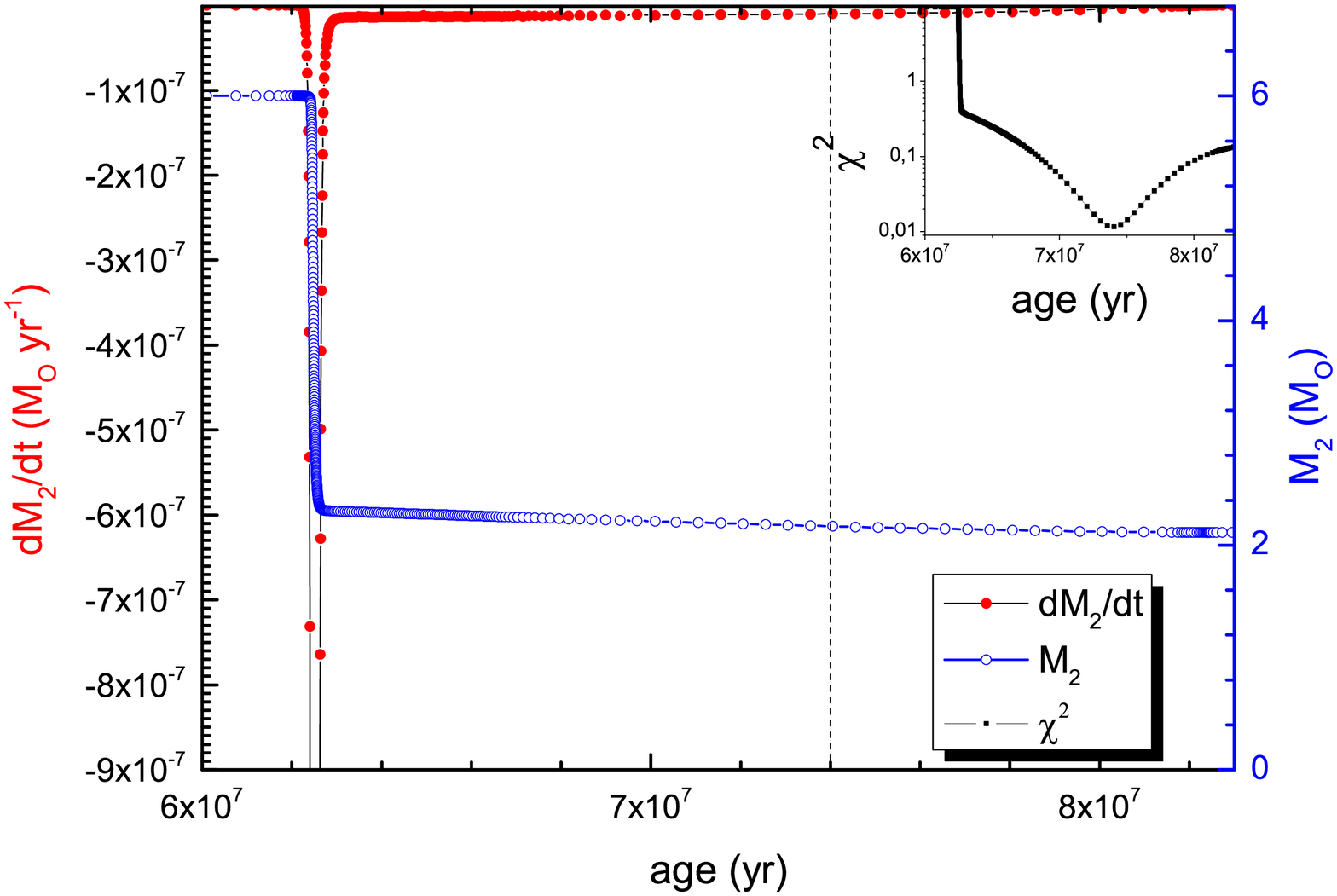}
   \caption{Best fit results for the evolution of DQ Vel with a gainer star under a synchronous rotation regime (upper panel) and, with a gainer star under
   a critical rotation regime (lower panel). The vertical dashed line represents the current stage of DQ Vel.}
              \label{Fig12}
    \end{figure}

\subsubsection{Gainer under a synchronous rotation regime}
For a gainer star rotating synchronously with the orbital motion, the best solution is obtained for a binary under a conservative evolution regime. 
In this case, the present state of DQ Vel is just inside a rapid mass transfer episode with a high mass transfer rate value of 
$\dot{ M_\mathrm{d}}=-8.6\times10^{-6}\,[\mathrm{M_{\odot}/yr}]$ (upper panel at Fig. 12). Such a high mass transfer rate, under a conservative 
scheme, would 
imply a significant change of
the orbital period. This change rate will be given by (Hilditch \cite{hilditch}), 
\begin{equation}
\frac{\dot{P}}{P}=\frac{3\dot{M_\mathrm{d}}(M_\mathrm{d}-M_\mathrm{g})}{M_\mathrm{d}M_\mathrm{g}}.
\end{equation}
Using the stellar masses and the orbital period obtained for a synchronous rotator gainer and the calculated value of the mass transfer rate, we 
find a 
change on the orbital period of 4 seconds per year. Considering a current accuracy of $10^{-6}$ on the determined orbital period, such a change
should be detected with
the current astronomical instruments. Furthermore, we can estimate from the model that DQ Vel has remained inside this high mass transfer episode
for
the last $\sim1.5\times10^{5}$ years. If we consider this mass transfer rate as a lower limit, we should expect at least a four minutes change 
on the binary
orbital period during the last 60 years. A comparision between the orbital period of $6.08337(13)\, \mathrm{d}$ found by van Houten 
(\cite{vanhouten}) with the current orbital period value, argues against this scenario.
\begin{table}
\caption{The two best theoretical solutions for the stellar and orbital parameters of DQ Vel. On the left side we list the parameters with a gainer star in synchronous
rotation regime and, in the right side are listed the parameters for a critical rotator gainer star.}     
\centering          
\begin{tabular}{l c r} 
\hline     
Gainer regime& Synchronous & Critical \\
\hline
 Parameter & Value&  Value \\ 
\hline                    
$\chi_{i,j}^{2}$& 0.0172 &0.0116   \\ 
Age [yr]& $6.78\times10^{7}$ & $7.40\times10^{7}$ \\  
Period [days]&6.120 & 6.057 \\ 
$M_\mathrm{d}\,[M_{\odot}]$& 2.351& 2.168 \\     
$\dot{M_\mathrm{d}}\,[\mathrm{M_{\odot}/yr}]$ & $-8.6\times10^{-6}$ & $-9.8\times10^{-9}$\\  
$log\,T_{d}$[K]& 3.89 & 3.94\\  
$log\,L_{d}\,[L_{\odot}]$& 2.40& 2.54\\
$R_{d}\,[R_{\odot}]$ & 8.69 & 8.22\\ 
$X_{cd}$& 0.06 & 0.07 \\ 
$Y_{cd}$ & 0.91 & 0.90\\ 
$log\,T_{g}$[K] & 4.32 & 4.31 \\  
$log\,L_{g}\,[L_{\odot}]$&3.38 &3.46  \\
$M_{g}\,[M_{\odot}]$ & 7.25 & 7.43 \\ 
$R_{g}\,[R_{\odot}]$ & 3.69 & 4.39 \\ 
$X_{cg}$ & 0.64 & 0.52\\ 
$Y_{cg}$ & 0.34 & 0.45\\ 
\hline     
\end{tabular}
\end{table}

\subsubsection{Gainer under a critical rotation regime}
The best evolutionary history of DQ Vel with the gainer under critical rotation (lower $\chi_{i,j}^{2}$) is achieved for a \textit{liberal} 
scenario under a weak tidal interaction regime (lower panel at Fig. 12). The theoretical stellar parameters obtained in this case are listed at the right column in Table 3. 
The better match between the observed and theoretical stellar properties is displayed at the lower panel in Figure 13. 
   \begin{figure}
   \centering
   \includegraphics[width=9.5cm]{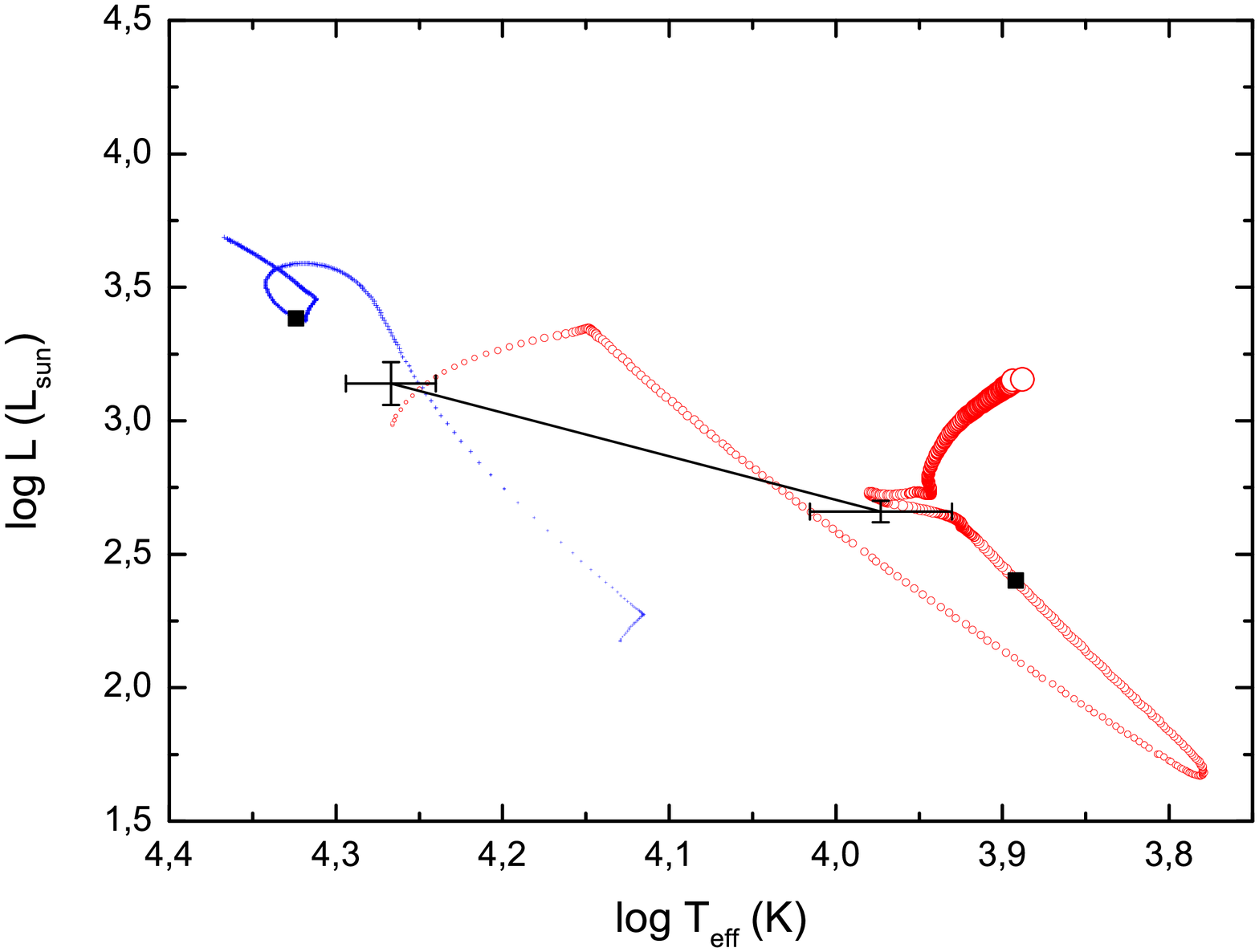}
    \includegraphics[width=9.2cm]{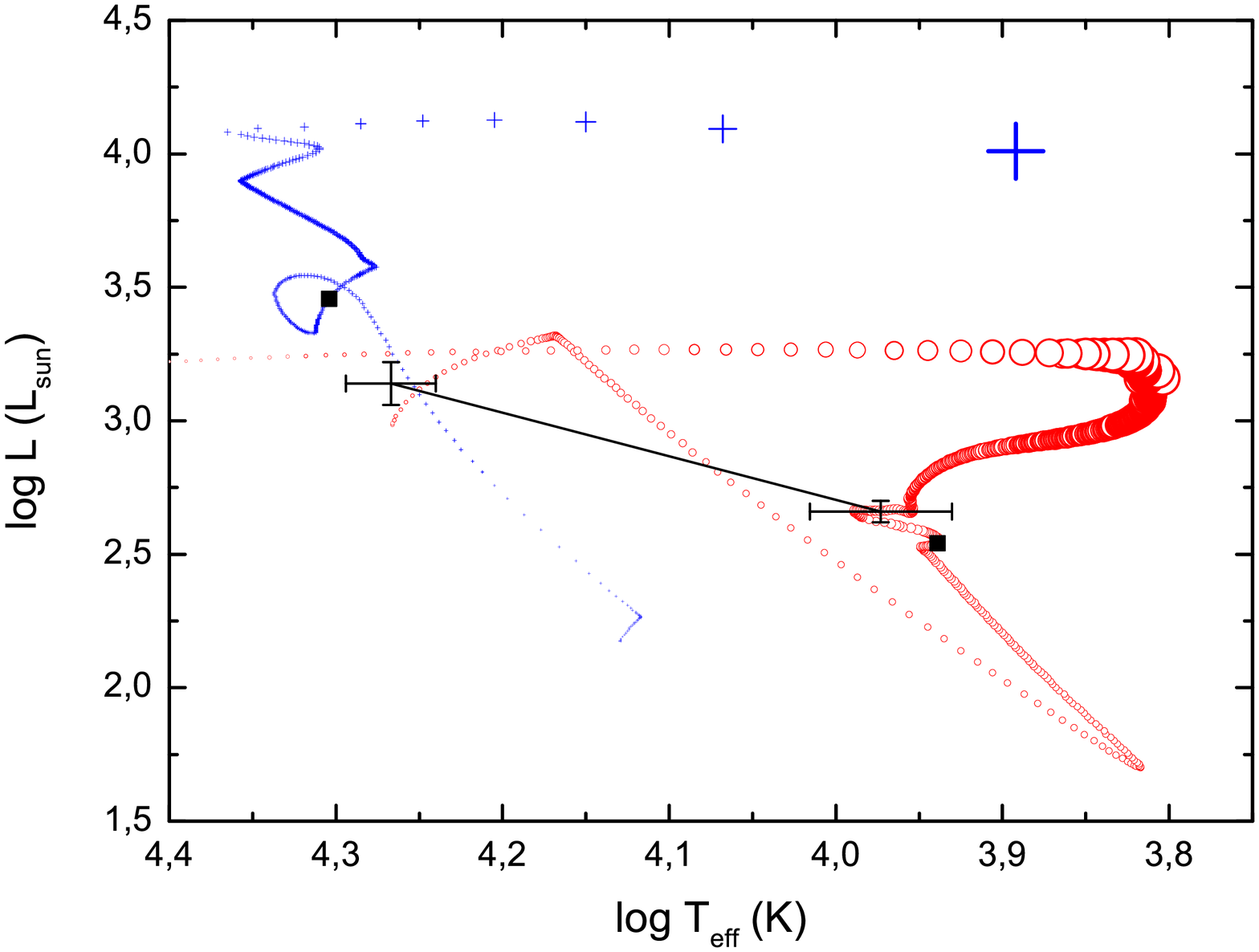}
   \caption{Evolutionary tracks for the DQ Vel components obtained from the best evolutionary model with a synchronous rotator gainer star (top panel) and, 
   a critical rotator gainer star (lower panel). Black squares and dots with error bars represent the current theoretical and the observed parameters respectively.}
              \label{Fig13}
    \end{figure}
Here, theoretical evolutionary tracks and physical parameters (black squares) for the donor and gainer are plotted together with the observed stellar 
parameters (black dots with error bars).
As a comparision in the upper panel of Figure 13 we show the theoretical evolutionary tracks and stellar parameters for the gainer and donor under a 
conservative evolution scenario. Higher discrepancies between the synthetic and observed parameters are found under this regime.\\
As can be seen at the lower panel in Figure 13 the observed gainer appears under-luminous and under-heated than the synthetic star. One possible explanation
for this might be that a fraction of the mass transferred from the donor has not been accreted by a rapidly rotating gainer and remains in form of a 
gaseous envelope and/or in form of an accretion disc around the more massive star.\\
To estimate the projected rotational velocity of the gainer ($v_\mathrm{g} \sin i$) we compare the helium profiles of the separated spectra with a grid of synthetic 
spectra at different rotational velocities. The best match is obtained for a synthetic spectrum with $v_\mathrm{g} \sin i = 120 \pm 25\,\mathrm{km\,s^{-1}}$. 
This value is around four times higher than the synchronous rotational velocity in agreement with our previous results for a gainer which is not 
rotating synchronously. However, it is important to consider here that a possible contamination in the helium lines due to the circumstellar material
may cause uncertainties in this velocity estimation.\\
The synchronization time scale for DQ Vel can be estimated by (Hilditch \cite{hilditch}),
\begin{equation}
 t_{sync}\approx 10^{4}\left(\frac{1+q}{2q}\right)^{2} P_{o}^{4} \,\,\,[\mathrm{years}] \approx 3.4 \times 10^{6}\,\, [\mathrm{years}]\,,
\end{equation}
where $q$ represents the system mass ratio. However, the evolutionary studies suggest that DQ Vel is an older system (see Figure 12) and 
thus we should expect that the stellar rotation periods have already been synchronized with the orbital period. A faster rotating gainer star 
can be explained then as a result of mass transfer processes in the system. The rotational parameter $F[(v_\mathrm{g} \sin i)_{obs}/(v_\mathrm{g} \sin i)_{sync}]\sim4$ 
for DQ Vel gainer is consistent with the F values of gainer for classical Algol systems with $P_{o}> 5\,\mathrm{d}$ given by 
Soydugan et al. \cite{soydugan2013} (see Figure 9).

\subsection{Accretion disc}
To check the possible formation of an accretion disc, we use the mass ratio and the current orbital separation to compute
the distance of closest approach (measured from the center of the gainer) of a stream coming from the inner Lagrange point $L_{1}$. 
From Lubow and Shu (\cite{lubow}), we obtain:
\begin{eqnarray}
 r_{min}&=&0.0488\,q^{-0.464}\,a \nonumber \\
 r_{min}&=&(0.084\pm0.011)\,a \,.  
\end{eqnarray}
The currently observed gainer radius $R_{g}=(0.135\pm0.05)\,a$ is a limiting case (inside the errors) to naturally form an accretion disc. 
To investigate if an accretion disc might have been formed in the past, we explore the evolutionary history of DQ Vel at different ages to determine
if the condition $R_{g}<r_{min}$ was achieved before. 
At present, this condition has not been fulfilled. However, the uncertainties in the $q$ and $R_{g}$ values do not allow to unambiguously reject a 
prior natural formation of a disc. In addition, the fact that the best light curve model is found for a semidetached configuration which includes an
extended accretion disc, argues against this fact. Furthermore, we can not exclude a different mechanism to form an accretion disc around the gainer. 
One possibility is that during a high mass 
transfer rate episode, the gainer was not able to accrete all the material coming from the donor, reaching a critical rotation where a fraction of
the transfered mass was accumulated close to the star. Once the mass transfer rate decreases the gainer may slow down and the accumulated matter 
spreads forming an accretion disc. The tidal interaction with the donor will continuosly break the rotational velocity of the gainer leading the 
material to move outwards creating an extended disc. These assumptions together with the lower mass transfer rate 
($-9.8\times10^{-9}\,\mathrm{M_{\odot}/yr}$) found for a non-syncronous gainer, seem to be a better 
representation of the current observed properties of DQ Vel. As a comparision, V393 Sco is a younger system just after a high mass transfer rate 
episode, where most likely the gainer is currently rotating critically with most of the transferred mass accumulated close to the gainer forming a
massive disc, whose thickness is higher in the inner part than in the outer rim of the disc (see Table 8 in Barr\'ia et al. \cite{barria}).

\subsection{Mass transfer rate}
Ibano\u{g}lu et al. (\cite{ibanoglu}) found significant Carbon deficiencies in the mass-gaining primaries of some classical Algol binaries. 
The authors measured the EWs of the C\,II\,4267 $\mathrm{\AA{}}$ line in 18 Algol primaries and compared them to the EW of single standard stars
having the same effective temperatures. The EWs of the Algols were significantly smaller than the standard stars. This result suggest that primary 
components of some Algol systems are C poor stars, due to the mixing processes in the primary atmospheres. Since the transferred material from the 
evolved donor star is C poor, when this is mixed with the original matter of the gainer the C amount appears to decrease (de Greve \& Sarna \cite{degreve}).
 Thus, Ibano\u{g}lu et al. (\cite{ibanoglu}) found a correlation between the EWs of C\,II\,4267 $\mathrm{\AA{}}$ line and the mass transfer rate. 
 As this latter increases the EW of C\,II\,4267 $\mathrm{\AA{}}$ decreases. Interestingly, in the Algols analysed by the authors the weakest line was 
 found in the DPV system AU Mon 
(originally classified as an Algol-type star) where the EW of the C\,II line was only 20 per cent that of a standard star with similar effective temperature. 
This result suggests that AU Mon has a higher mass transfer rate value compared to the classical Algol systems studied by the authors.\\
To obtain an independent estimation of the mass transfer rate in DQ Vel, we measured the EW/$\lambda$ for the C\,II\,4267 $\mathrm{\AA{}}$ line at five
donor substracted spectra collected at orbital phase $\phi_{o}=0.5$ when the donor star is being eclipsed. Using the correlation EWs/$\dot{M}$ for
the Algols studied by Ibano\u{g}lu et al. (\cite{ibanoglu}) we measured a mean value of the 
EW/$\lambda$ of $-4.7\pm0.3$ and thus we estimate a mass transfer rate around $\dot{M}=-5\times10^{-8}\,\mathrm{M_{\odot}/yr}$. 
This mass transfer rate is lower than the value predicted from the evolutionary analysis of DQ Vel with the gainer star in
synchronous rotation (see Section 3.4.1 and Table 3), but is closer to the mass transfer rate estimated for a gainer in critical rotation
($-9.8\times10^{-9}\,\mathrm{M_{\odot}/yr}$). 
Considering that the gainer star in DQ Vel is rapidly rotating (120 $\mathrm{km\,s^{-1}}$) but not critical (see Section 3.4.2), we suggest that 
$\dot{M}\sim10^{-8}\,\mathrm{M_{\odot}/yr}$ is a better estimation of the mass transfer rate of this system. \\
We compare the observed EW for the C\,II\,4267 $\mathrm{\AA{}}$ line with that of a synthetic star with the same effective temperature and 
luminosity class of the DQ Vel gainer. The observed EW of the C\,II\,4267 $\mathrm{\AA{}}$ line is around 40 per cent of the EW measured in the synthetic star. 
Comparing this result with that of AU Mon (see text above), we suggest that C in the DQ Vel gainer atmosphere is not as diluted as in AU Mon gainer 
and, thus the mass transfer rate should be higher in AU Mon than in DQ Vel. 

 \section{Conclusions}
We have analised a series of optical spectra of DQ Vel during much of its long-term cycle. We have found remarkable profile variations in
the donor-subtracted and in the difference spectra at similar orbital phases, suggesting that an additional source is responsible for these variabilities.
The EWs of the central absorption measured in Balmer and helium lines seems to be modulated with the long cycle. During the minimum of the long-term
variability we observe a strenghtening in the EWs in all analysed spectral features. Moreover, during the long-term maximum all profiles appear 
weakened. This pattern is most remarkable at the main eclipse when the gainer star and the disc are being eclipsed by the donor star. The constancy of
the orbital light curve during the long cycle, suggests that this variability pattern is not related to changes in the disc properties. Furthermore, the additional source is not 
eclipsed and thus seems to be restricted to some region above/below the orbital plane. We suggest that the spectral variabilities observed
during the long-term phase are better represented by an extra line
emission produced during the maximum brightness of the system which decreases the EWs at the long-term maximum. A similar behaviour was also observed
in the DPV system V393 Sco.\\
Discrete absorption components (DACs) were found at three donor subtracted spectra in the O\,I\,7773 $\mathrm{\AA{}}$ absorption line. DACs were 
detected at the orbital phases $\phi_{o}=0.32, 0.48$ and $0.65$ together with blue enhanced absorption wings observed at the main absorption profile. 
Enhanced absorption wings and DACs seems to be correlated. DACs are only present when a blue or red enhanced absorption wing is observed.
We suggest that DACs and enhanced absorption wings can be associated to a clumpy wind coming from the accretion disc.\\
The stellar parameters of DQ Vel obtained for a gainer star in synchronous and critical rotation were compared with a grid of theoretical 
evolutionary tracks. These tracks were computed under two different evolutionary scenarios: a conservative and a non-conservative (\textit{liberal}) 
regime. A better representation of the current observed parameters of DQ Vel was found for a non-conservative evolutionary case, where mass has been lost
for the system at some stage of the binary history. The best non-conservative evolutionary model for DQ Vel was found with the gainer star in critical
rotation and the donor transfering mass 
at a rate of $-9.8\times10^{-9}\,\mathrm{M_{\odot}/yr}$.\\
In an independent way, we measured the EW of the C\,II\,4267 $\mathrm{\AA{}}$ line and estimate a higher mass transfer rate of 
$-5\times10^{-8}\,\mathrm{M_{\odot}/yr}$. 
The EW of the C\,II\,4267 $\mathrm{\AA{}}$ line is around 40 per cent lower than that of a synthetic star with the same effective temperature
and luminosy class of the DQ Vel gainer. 
After comparing the gainer helium lines in the separated spectra with a grid of synthetic spectra we estimate
a gainer rotational velocity of $v_\mathrm{g} \sin i = 120\, \mathrm{km\,s^{-1}}$, which is around four times higher than the synchronous rotational 
velocity. This result however considers that the helium lines used to estimate $v_\mathrm{g} \sin i$ comes from the photosphere 
of the gainer.\\
After comparing the evolutionary stages of DQ Vel and V393 Sco, we have found that DQ Vel seems to be an older system with a lower mass transfer rate. 
DQ Vel and V393 Sco show similar stellar parameters but the geometrical and physical properties
of their accretion discs are different. The gainer star in DQ Vel seems to be a rapidly rotating star while V393 Sco gainer is a critical rotator.
We suggest that in the older system DQ Vel, the material transfered from the donor has had more time to spread out around the gainer forming an 
extensive accretion disc. On the other hand, the massive disc observed in V393 Sco can be explained by a younger system with a higher mass transfer
rate where the critical rotator gainer might not have had enough time to accrete all the matter coming from the donor and thus 
to form a massive accretion disc.

\begin{acknowledgements}
We thanks to Sandro Villanova, Guisella Moreno, Andr\'e-Nicolas Chen\'e, Matias Jones and Paz Bluhm for carry out part of the observations used in this work. 
D. B. acknowledges support by the Chilean CONICYT PhD grant. R.E.M. acknowledges support from FONDECYT grant 1110347. D. B. and R.E.M. acknowledge support from the BASAL 
Centro de Astrof\'isica y Tecnologias Afines PFB-06/2007.
\end{acknowledgements}

\end{document}